\documentclass[%
 reprint,
 amsmath,amssymb,
 aps,
]{revtex4-2}

\usepackage{comment}

\usepackage{graphicx}
\usepackage{dcolumn}
\usepackage{bm}
\usepackage{booktabs}

\usepackage{footnote}
\usepackage{url}
\usepackage{hyperref}
\usepackage{multirow}
\usepackage{braket}
\usepackage{ulem}
\usepackage{xcolor}

\begin{document}

\preprint{APS/123-QED}

\title{A hardware efficient quantum residual neural network without post-selection}

\author{Amena Khatun$^{\ast}$\textsuperscript{1}}
\author{Akib Karim$^{\ast}$\textsuperscript{1}}
\thanks{These authors contributed equally}

\thanks{Corresponding author: amena.khatun@csiro.au}

\author{Muhammad Usman\textsuperscript{1,2}}

\affiliation{%
\textsuperscript{1}Quantum Systems, Data61, CSIRO, Australia\\
\textsuperscript{2}School of Physics, The University of Melbourne, Victoria, Australia}%

\begin{abstract}
We propose a hardware efficient quantum residual neural network which implements residual connections through a deterministic mixture of the identity operation and variational unitaries, enabling fully differentiable training. In contrast to the previous implementation of residual connections, our architecture avoids post-selection while preserving residual learning. Furthermore, we highlight circuit constructions where barren plateaus could be mitigated, which are considered as a major limitation of variational quantum learning models. In order to show the working of our model, we report its application to image classification tasks by training it for MNIST, CIFAR, and SARFish datasets, achieving accuracies of 99\% and 80\% for binary and multi-class classifications, respectively. These accuracies are comparable to previously achieved from the standard variational models, however our model requires 10x fewer gates making it better suited for resource constraint near-term quantum processors. In addition to high accuracies, the proposed architecture also demonstrates adversarial robustness  which is another desirable parameter for quantum machine learning models. Overall our architecture offers a new pathway for developing accurate, robust, trainable and hardware efficient quantum machine learning models.

\end{abstract}

\maketitle

\section{Introduction}
\label{intro}

Quantum machine learning (QML) has been touted as one of the most promising applications for near-term quantum computers. In the last few years, rapid progress in quantum hardware and software has catalyzed the development of quantum analogues of many classical machine learning algorithms, including classifiers~\cite{schuld2015introduction, farhi2018classification}, kernel methods~\cite{havlivcek2019supervised, schuld2019quantum}, and neural networks, with early demonstrations spanning image classification~\cite{west2023benchmarking, khatun2024quantum}, image generation~\cite{khatun2025quantum}, pattern recognition~\cite{schuld2014quantum}, and signal processing~\cite{wu2023radio}. Among these approaches, quantum variational classifiers (QVCs) have emerged as dominant framework for QML implementation on noisy intermediate-scale quantum (NISQ) devices. In this paradigm, classical data are embedded into quantum states, processed through parameterized quantum gates, and measured to produce task-specific outputs. QVCs offer differentiability and enable end-to-end training using classical optimization techniques. Several studies have reported accuracies from QVC models comparable to classical neural networks albeit for only proof-of-concepts examples ~\cite{west2023benchmarking, abbas2021power, huang2021power, havlivcek2019supervised}, including some studies predicting adversarial robustness to classical attacks~\cite{khatun2024quantum, 41z8-d3h9, west2023benchmarking, west2023towards, wu2023radio}. However, challenges remain in the scalability and generalization of QVC methods~\cite{larocca2025barren, thanasilp2023subtleties, heyraud2023efficient}, in particular limitations arising from the presence of barren plateaus, classical simulability and deep circuits incompatible with the near-term quantum devices. An end-to-end differentiable and trainable quantum machine learning framework which is also hardware efficient remains an open research problem.

In this work, we propose a QML architecture that implements residual connections explicitly via a deterministic mixture of the identity operation and variational unitaries and combines the benefits of both QResNet and density quantum machine learning. We use ancilla-controlled residual blocks without post-selection to allow for non-linear state concentration and efficient backpropagation. Our approach overcomes limitations such as probabilistic execution, limited compatibility with gradient based optimization and the lack of generality pertaining to previously proposed models such as density based technique~\cite{Coyle2025}, or alternative residual-style architectures~\cite{dibrita2025resq, yang2025quantum}. We demonstrate trainability on MNIST and CIFAR-2 datasets with accuracies at par with the standard QVC techniques but significantly fewer gates, offering a hardware efficient pathway compatible for near-term quantum devices. Furthermore, we benchmark the adversarial robustness of our model exhibiting that our model retains high accuracy under black-box setting when the adversarial attacks are transferred from classical models.


\section{Literature Background}
While QML has demonstrated promising capabilities, the trainability of variational quantum models remains a fundamental challenge. Several approaches have been proposed to address this limitation. Here we discuss only the approaches that address trainability limitations in variational quantum circuits, with a focus on methods most relevant to architectural and formulation-based strategies related to residual quantum models. Ref.~\cite{kieferova2021quantum} introduces loss functions based on Rényi divergence that modify gradient concentration behavior, and demonstrates that under specific conditions such formulations can avoid the exponential suppression of gradients associated with barren plateaus. Similarly, Ref.~\cite{yao2025avoiding} proposes an entanglement based circuit construction using auxiliary control qubits to mitigate barren plateaus by transforming the circuit, preventing the circuit from approaching highly random transformations that lead to vanishing gradients. However, this approach does not provide an explicit architectural mechanism for maintaining gradient propagation across successive variational layers as circuit depth increases.

Ref.~\cite{PhysRevApplied.23.044046} introduces a residual framework through coherent combinations of identity and variational transformation. However, this formulation relies on post selection for implementing non unitary operations, resulting in probabilistic state preparation and measurement. While the work includes analysis of gradient variance and barren plateau behavior, an explicit formulation of gradient propagation through the post-selected non-unitary residual construction is not presented, making its integration with standard gradient-based optimization less clear. Density-based approach has been proposed in Ref.~\cite{Coyle2025}, where coherent superposition is replaced by probabilistic mixtures of variational layers within a density matrix framework. This formulation enables efficient gradient evaluation and avoids the need for post-selection. However, it fundamentally alters the underlying mechanism by removing coherent interference between identity and transformed states, and does not provide a residual construction that explicitly regulates gradient propagation across successive transformations. Residual learning has also been introduced in analog quantum computing through continuous-time Hamiltonian evolution~\cite{dibrita2025resq} that is not directly compatible with standard gate-based circuit models. Similarly, attention-based residual mechanisms have been proposed within quantum neural network architectures~\cite{yang2025quantum}. While these formulations incorporate residual connections into specific model designs, an explicit framework ensuring end-to-end differentiability for gradient-based optimization is not established. In addition, a trainable parameterization that enables continuous control over the contribution of residual transformations across layers is not defined, limiting the ability to regulate information flow and gradient behavior in deeper circuits.

\section{Quantum Residual Neural Network}
\label{method}
\begin{figure*}
\begin{center}
\includegraphics[width=1.0\linewidth]{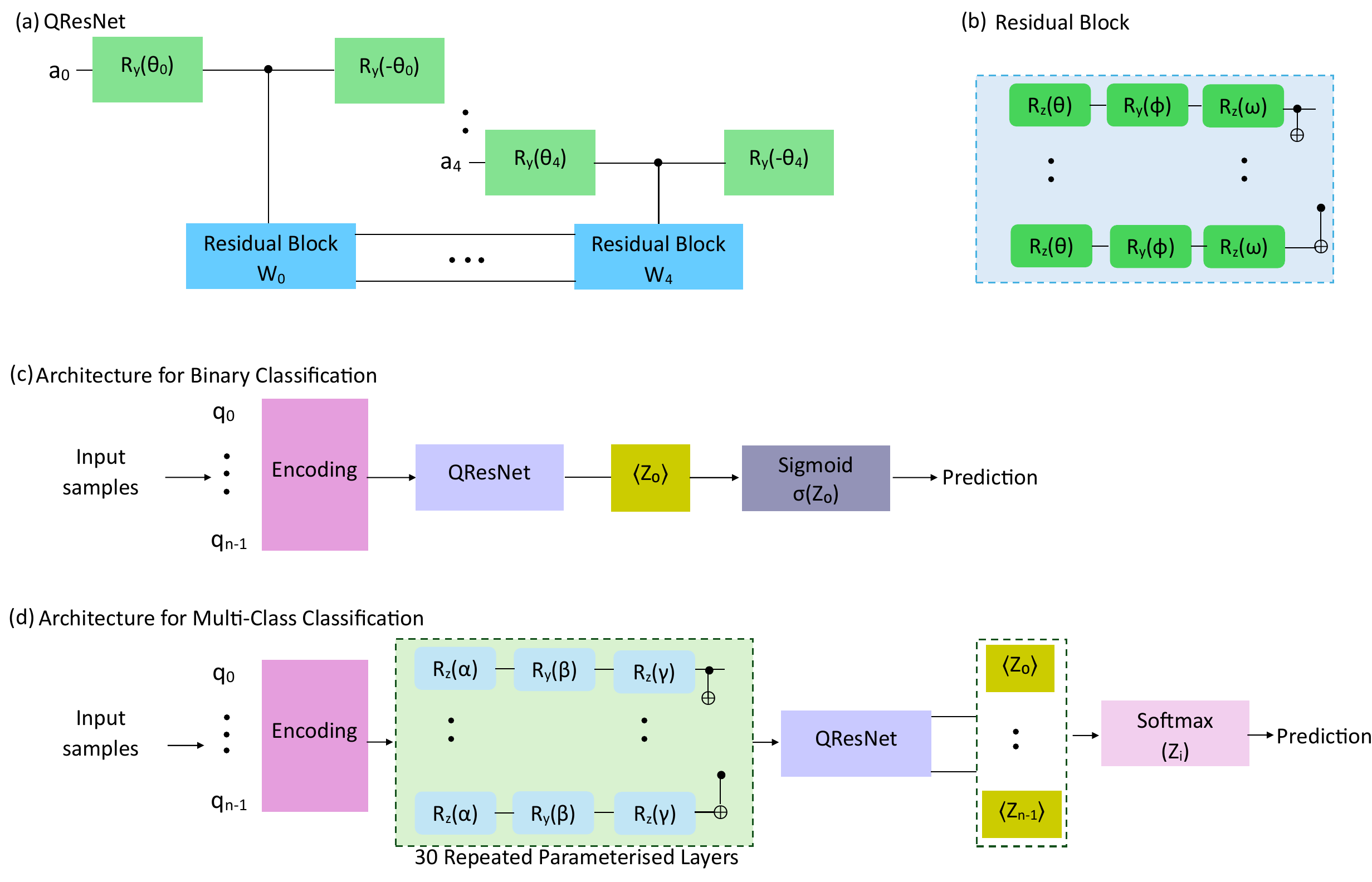}
\end{center}

\caption{\textbf{Overview of the proposed QResNet architecture.} 
(a) Classical input data are amplitude encoded onto the data qubits $q_0,\ldots,q_{n-1}$. A sequence of $L$ ancilla-controlled residual blocks is then applied. In the $\ell$-th block, ancilla qubit $a_\ell$ is prepared in a superposition state via $R_Y(\theta_\ell)$, where $\ell$ indexes the residual blocks (as illustrated in panel (a), e.g., $\theta_0, \theta_4$). The ancilla controls the application of the variational unitary $W_\ell$ on the data qubits and is subsequently unprepared using $R_Y(-\theta_\ell)$. This mechanism implements residual connections through an ancilla-controlled probabilistic mixture of the identity operation and variational unitaries. (b) Each residual block consists of parameterized single-qubit rotations $R_z(\theta)$, $R_y(\phi)$, and $R_z(\omega)$ applied to each data qubit, followed by entangling gates between neighbouring qubits. In this work, five residual blocks are employed. (c) For binary classification, the expectation value $\langle Z_0 \rangle$ is rescaled by the factor $\prod_{\ell=1}^{L} (1 + \beta_{\ell}^2)$ to produce the normalized logit. The final prediction is obtained via the sigmoid activation function $\sigma(z_0)=\frac{1}{1+e^{-z_0}}$. (d) For multi-class classification, additional parameterized quantum variational circuit (QVC), composed of 30 repeated layers of single-qubit rotations $R_z(\alpha)$, $R_y(\beta)$, and $R_z(\gamma)$ together with entangling gates, is applied prior to the residual blocks to enhance expressivity. Expectation values measured on all data qubits produce logits $z_i=\langle Z_i\rangle$ for $i=0,\ldots,n-1$, and the final class probabilities are obtained using the softmax function $\mathrm{softmax}(z_i)=\frac{e^{z_i}}{\sum_{j=0}^{n-1} e^{z_j}}$.}
\label{fig:QResNet}
\end{figure*}

Our QResNet implements the concept of skip-connections into quantum variational circuit by ancilla-controlled unitaries. The overview of QResNet is illustrated in Figure \ref{fig:QResNet}. The concept of skip connection is inspired by classical residual networks, where the shortcut paths stabilize optimization and mitigate vanishing gradients by allowing information to bypass non-linear transformations. In quantum implementation, residual connections are enabled by ancilla qubits that control whether a variational block acts on the data qubits. Tracing out the ancilla produces a quantum channel that probabilistically mixes the identity and unitary operation by a parameterized transformation. We use amplitude encoding \cite{larose2020robust} to map the classical data onto a quantum state, where each component of the classical data corresponds to the amplitude of the quantum state. For a classical dataset $x = (x_1, x_2,....,x_N)$, where $x_i$ is a normalized value, the quantum state can be written as:
\begin{equation}
    |\psi\rangle = \sum_{i=0}^{N-1} x_i |i\rangle,
\end{equation}
where $|i\rangle$ are the computational basis states and $N = {2^n}$ (where $n$ is the number of qubits).

Each ancilla-controlled residual block uses a single ancilla qubit to determine whether the data qubit are transformed by a variational unitary or remain unchanged. We employ a single-qubit $R_Y$ rotation for superposition as:
\begin{equation}
R_Y(\theta) |0\rangle = \cos\left(\tfrac{\theta}{2}\right)|0\rangle + \sin\left(\tfrac{\theta}{2}\right)|1\rangle
\end{equation}
Here, we consider a sequence of ancilla-controlled residual blocks indexed by $\ell \in \{0, \ldots, L-1\}$, as illustrated in Figure~\ref{fig:QResNet}(a). Each residual block is associated with an ancilla qubit $a_\ell$, a variational unitary $W_\ell(\vartheta_\ell)$, and a residual strength parameter $\beta_\ell$ that controls the relative contribution of the identity and the variational transformation. The preparation angle for the ancilla qubit in the $\ell$-th block is defined as $\theta_{l} = 2 \arctan\left( \lvert \beta_{l} \rvert \right)$ which gives amplitude $c_{l} = \frac{1}{\sqrt{1 + \lvert \beta_{l} \rvert^{2}}}$, and $s_{l} = \frac{|\beta_l|}{\sqrt{1 + \lvert \beta_{l} \rvert^{2}}}$. A phase shift of $\pi$ is applied in $\beta_{l} < 0$, this phase is undone before the ancilla is uncomputed, hence, the final residual map depends only on  $|\beta_{l}|$. The ancilla qubit is therefore prepared in a coherent superposition of the computational basis states. A controlled unitary is then applied, where the identity operation acts on the data qubits conditioned on the ancilla being in the $|0\rangle $ state, while the variational unitary acts conditioned on the ancilla being in the $|1\rangle $ state. The variational transformation $W_l(\vartheta_l)$ is parameterized single-qubit rotations on all data qubits followed by entangling gates. This is given by:
\begin{equation}
    W_l(\vartheta_l) = \prod_k A_k\prod_j e^{-i Z_j \omega_j}e^{-i Y_j \phi_j}e^{-i Z_j \theta_j},
\end{equation}
where $Y_j,Z_j$ represent Pauli matrices on the $j$th qubit; each qubit has trainable angles $\theta_j, \phi_j,\omega_j$ where we group all parameters in the layer as $\vartheta_l$; and $A_k$ are the entangling gates over all adjacent pairs of qubits $k$, in our case they are CNOTs.

Details of the residual circuit is illustrated in the Appendix (see Figure~\ref{fig:circuit}). This ansatz offers high expressivity to capture local and nonlocal correlations. After the controlled operation, the inverse rotation $R_Y(-\theta_l)$ is applied to the ancilla qubit. Because the controlled-$  W_l  $ gate entangles the ancilla with the data register, tracing over the ancilla yields a quantum channel on the data qubits.

This prepare, control and uncompute sequence is a related to the LCU technique. In LCU, an ancilla prepared in a superposition coherently selects between different unitaries, and post-selecting the ancilla outcome implements a linear combination of those unitaries on the data qubit. In Ref. \cite{PhysRevApplied.23.044046}, the ancilla amplitudes are chosen such that post-selection yields a residual map proportional to $I + \beta_l W_l$. In our approach, we adopt a different ancilla preparation rule. Tracing out the ancilla after the prepare–control–uncompute sequence yields a completely positive, trace-preserving channel on the data register. Let $  \rho  $ denote the density matrix of the input state to the $  \ell  $-th residual block,

\begin{equation}
\mathcal{E}_l(\rho)
=
\frac{1}{1 + |\beta_l|^2}\,\rho
+
\frac{|\beta_l|^2}{1 + |\beta_l|^2}\,
W_l(\vartheta_l)\,\rho\,W_l(\vartheta_l)^\dagger,
\label{eq:4}
\end{equation}
a classical probabilistic mixture of the identity channel and the unitary conjugation by $  W_l(\vartheta_l)  $. We note that this means the residual connections are modified from the original proposal so that we have classical mixtures between layers rather than a coherent superposition between layers.
The original QResNet formulation \cite{PhysRevApplied.23.044046}, prepares each ancilla as $\sqrt{1 - \beta_l} |0\rangle + \sqrt{\beta_l} |1\rangle$ and, after post-selection on $|0\rangle$, yields a coherent map proportional to $(1 - \beta_l)I + \beta_l W_l(\vartheta_l)$.
In contrast, we deliberately adopt $\theta_l = 2 \arctan(|\beta_l|)$ (with a conditional phase shift of $\pi$ for $\beta_l < 0$) and never post-select. This produces amplitudes $c_l = 1/\sqrt{1 + |\beta_l|^2}$ and $s_l = |\beta_l|/\sqrt{1 + |\beta_l|^2}$. The expectation value $\langle Z_0 \rangle$ is automatically scaled by the multiplicative factor $(1 + |\beta_l|^2)$. 

The resulting deterministic surrogate
\begin{equation}
f(x) = \left[ \prod_{\ell=1}^L (1 + \beta_\ell^2) \right] \langle Z_0 \rangle ~\label{eq:fx}
\end{equation}
is fully differentiable with respect to both variational angles and residual strengths. This choice also gives $\beta_l$ a clear physical meaning as the relative strength of the variational unitary versus the identity channel, with transparent limits: $\beta_l \to 0$ bypasses the block, while $|\beta_l| \to 1$ applies an equal mixture $\frac{1}{2}(I + W_l)$. The parameterization was specifically engineered to eliminate post-selection while preserving the residual structure and barren-plateau mitigation. 

Thus each block realizes a weighted combination of the identity and the variational transformation, with weights determined directly by the ancilla amplitudes. Although no post-selection is performed, it is instructive to define the notional probability that the ancilla would be found in  $|0\rangle $ if it were measured:
\begin{equation}
p_l(\psi_{\mathrm{in}}) \;=\; \frac{1}{(1+|\beta_l|^2)^2}
\Big(1 + |\beta_l|^4 + 2|\beta_l|^2\,\operatorname{Re}\langle \psi_{\mathrm{in}}|W_l(\vartheta_l)|\psi_{\mathrm{in}}\rangle\Big),
\end{equation}
where $\operatorname{Re}$ denotes the real part. For a circuit of $L$ residual blocks, the transformation can be expressed as,
\begin{equation}
\rho_L = \mathcal{E}_L \circ \cdots \circ \mathcal{E}_1 \bigl(\rho(x)\bigr),
\qquad
\rho(x) = |\psi(x)\rangle\langle\psi(x)|
\end{equation}
and total probability that the circuit succeeds across all blocks is the product of the individual success probabilities,
\begin{equation}
\prod_{l=1}^{L} p_l
\end{equation}

\begin{figure}
    \centering
    \flushleft{(a)}
    \includegraphics[width=\columnwidth]{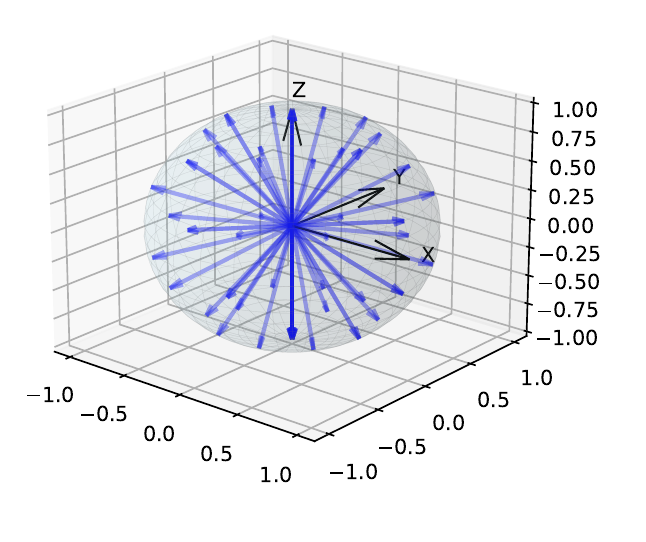}
    \vspace{-40pt}
    \flushleft{(b)}
    \includegraphics[width=\columnwidth]{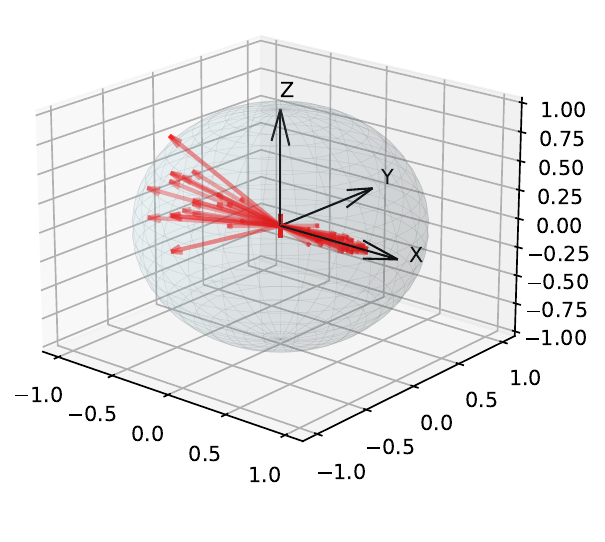}
    \vspace{-20pt}
    \caption{\textbf{Effect of non-postselected QResNet on Bloch Sphere.} Equally spaced one qubit statevectors are given in a). One layer QResNet with unitary $W=X$ and an ancilla preparation angle $\beta = (\frac{\pi}{2}-0.05)$. The output states were normalized and plotted in b). Unitary rotations preserve distances between input states, whereas QResNet allows for states to concentrate.}
    \label{fig:bloch}
\end{figure}

The effect of this on a quantum state is shown in Figure~\ref{fig:bloch} for a one qubit state. Evenly spaced states are shown on the Bloch Sphere to represent input states to the layer in Figure~\ref{fig:bloch} (a). While QVC is limited to training unitary operations which, in this case, correspond to rotating the state on the Bloch Sphere, our QResNet variation can cause concentration along an axis, which is a non-unitary effect. For Figure~\ref{fig:bloch}, we use Pauli $X$ as the unitary given as $W_l$ in Equation~\ref{eq:4}. A post-selected coherent LCU would select only the states in the $+x$ direction, while we retain the states in the $-x$ direction as well. This non-unitary effect allows for concentration, while the lack of post-selection means we do not concentrate to one axis but bifurcate to the antiparallel axis.

Our approach makes end-to-end trainable QResNet by avoiding post-selection entirely. In \cite{PhysRevApplied.23.044046}, LCU rely on probabilistic post-selection approach: only the circuit executions in which the ancilla is measured in state $|0\rangle$, while all other outcomes are discarded. This probabilistic acceptance decays exponentially with circuit depth. The deeper the network, the smaller the chance of obtaining an all $|0\rangle$ outcome across ancillas. On real quantum hardware, this would require an exponential number of repetitions to collect valid samples. More importantly, this approach is a sample-and-discard mechanism that is not differentiable. Gradients cannot propagate through stochastic measurement and rejection processes, since derivatives cannot be taken through a discrete sample-and-discard step. This breaks the computational graph and prevents the use of standard optimization methods. As a result, the canonical LCU approach is not suitable for gradient-based training.

To train a QML model, the parameters in variational quantum algorithms must be optimized to minimize a task-specific loss function (e.g., classification loss). This optimization requires gradients of the loss with respect to circuit parameters. We address these issues by never sampling the ancilla. The circuit returns the expectation value $\langle Z_0 \rangle$, while residual strengths $\beta_l$ provide a deterministic scaling that captures the effect of residual connections without post-selection. The ancilla-controlled residual construction introduces a parameter-dependent normalization factor arising from the probabilistic mixture of identity and variational unitaries. Rather than relying on post-selection of measurement outcomes, the resulting expectation value is deterministically rescaled by a normalization term controlled by the residual strengths $\beta_\ell$. Since both $\langle Z_0 \rangle$ and each $p_l$ are expectation values of observables in a differentiable circuit, $f(x)$ is a smooth function of all variational angles $\vartheta_l$ and residual strengths $\beta_l$. This retains the semantics of conditioning on success while enabling standard gradient-based optimization. The gradient of $f(x)$ can be computed by applying the product rule,
\begin{equation}
\nabla f(x)
= \left(\prod_{\ell=1}^{L}(1+\beta_\ell^2)\right)\nabla\langle Z_0\rangle
+ \langle Z_0\rangle
\nabla\left(\prod_{\ell=1}^{L}(1+\beta_\ell^2)\right)
\end{equation}
where $p_l$ denotes the success probability of the $l$-th ancilla, and
\begin{equation}
    \nabla\left(\prod_{\ell=1}^{L}(1+\beta_\ell^2)\right)
=
\left(\prod_{\ell=1}^{L}(1+\beta_\ell^2)\right)
\sum_{\ell=1}^{L}
\frac{2\beta_\ell}{1+\beta_\ell^2}\nabla\beta_\ell.
\end{equation}
This confirms that $f(x)$ is differentiable with respect to both the variational parameters and residual strengths. 

The strategy of measuring a single data qubit is sufficient for binary classification as a scalar logit can distinguish between two classes. However, a single observable expectation provides only one decision boundary and therefore cannot represent more than two class outcomes. Multi-class classification requires measuring the expectation values of all data qubits simultaneously,
\begin{equation}
(\langle Z_0\rangle, \langle Z_1\rangle, \ldots, \langle Z_{n_q-1}\rangle)
\end{equation}
This vector of expectation values provides multiple output channels, one per data qubit. As in the binary case, the ancilla qubits success probabilities are computed as expectation values and used to form a deterministic scaling
factor. The final network output is expressed as,
\begin{equation}
f(x)
=
\left(\prod_{l=1}^{L}(1+\beta_l^2)\right)
\big(\langle Z_0\rangle,\ldots,\langle Z_{n_q-1}\rangle\big).
\end{equation}


During training, the network parameters are updated using gradient-based optimization, where the training objective is defined by the cross-entropy loss. Cross-entropy is the standard objective for multi-class classification tasks, as it penalizes the discrepancy between the predicted logit vector ${f}(x)$ and the true class label. For an input $x$ with label $y \in \{0,\ldots,n_q-1\}$, the cross-entropy objective is,
\begin{equation}
\mathcal{L}(x,y) \;=\; - \log \frac{\exp(f_y(x))}{\sum_{j=0}^{n_q-1} \exp(f_j(x))},
\end{equation}
where the denominator normalizes the logits into a valid probability distribution over all classes, and the numerator selects the probability assigned to the correct class $y$. Minimizing $\mathcal{L}(x,y)$ therefore encourages the model to assign high probability to the correct class and low probability to all others.

In contrast to the fixed residual strengths used in the theoretical formulation of QResNet \cite{PhysRevApplied.23.044046}, in our approach,
we treat each $\beta_l$ as a trainable parameter. 
This choice of trainable $\beta_l$ provides several advantages.  It allows the network to adaptively regulate the balance between identity and transformation across layers, analogous to skip-connections in classical residual networks. A fixed $\beta$ enforces the same residual weighting in every block, regardless of depth or data distribution, whereas trainable $\beta_l$ values enable different layers to specialize. For example, optimization may drive some $  \beta_l \approx 0  $ (effectively bypassing those blocks) while pushing other $  \beta_l  $ closer to $  \pm 1  $ (emphasizing the action of the variational unitary). This adaptive behavior is observed after training on the datasets, such as for the MNIST binary classification task: the learned residual strengths converge to $  \beta_0 \approx 0.999  $, $  \beta_1 \approx 0.999  $, $  \beta_2 \approx 0.999  $, $  \beta_3 \approx 0  $, $  \beta_4 \approx -0.999  $. This indicates that the network learns to bypass the fourth residual block ($\beta_3 \approx 0$, applying only the identity operation) while keeping the remaining four blocks at full strength ($|\beta_l| \approx 1$, applying the full variational unitary $W_l$). Within our deterministic probability-scaling framework, both the variational angles $\vartheta_l$ and the residual strengths $\beta_l$ remain fully differentiable, ensuring compatibility with gradient-based optimization.


The role of $\beta_l$ can be further understood by examining the limiting behavior of the effective map defined in Equation (4). In the limit $\beta_l \to 0$, the block reduces to the identity,
\begin{equation}
\lim_{\beta_l \to 0} \mathcal{E}_l(\rho) = \rho
\end{equation}
so the layer is effectively bypassed. 
In contrast, when $|\beta_l| \to 1$, the block approaches an equal mixture of the identity and the variational unitary,
\begin{equation}
\lim_{|\beta_l| \to 1} \mathcal{E}_l(\rho)
=
\frac{1}{2}\Big(\rho + W_l(\vartheta_l)\,\rho\,W_l(\vartheta_l)^\dagger\Big)
\end{equation}
Thus, by optimizing $\beta_l$, the network can interpolate smoothly between bypassing a block and applying it with maximum residual strength. 
This flexibility stands in contrast to the fixed choice of $\beta$ (e.g., $\beta=0.5$) used in the original formulation~\cite{PhysRevApplied.23.044046}, which enforces uniform residual weighting across all layers. In the proposed framework, the trainable residual strengths provide explicit control over the contribution of each transformation, enabling adaptive regulation of information flow and contributing to stable gradient behavior in deeper circuits.

\section{Quantum Variational Circuit}
\label{qvc}
We found that a depth of five QResNet residual blocks is sufficient to achieve high performance on binary classification tasks. However, multi-class classification requires greater expressive power to capture complex decision boundaries. As quantum hardware are still being developed with a limited number of qubits, high error rates and decoherence, at this stage, we conducted all experimental simulations using an open-source software framework. To extend the model beyond this setting, we introduce additional variational quantum layers prior to the residual blocks, as illustrated in Fig.~\ref{fig:QResNet}(d). Each layer consists of parameterized single-qubit rotations applied to all data qubits, followed by entangling CNOT operations. This provides a standard variational circuit structure operating directly on the encoded quantum state before the application of residual transformations. This framework establishes a direct pathway for combining the proposed QResNet formulation with conventional variational quantum circuit designs. The variational layers can be incorporated without modifying the residual mechanism, and the resulting circuit remains fully differentiable with respect to all parameters. These additional variational layers do not affect the trainability of the model. As shown in Section \ref{scalability}
 and Appendix \ref{appen-A}, the residual contribution can help maintain gradient signal for the residual strength parameters in the overall objective even when additional variational components are introduced. This provides a mechanism for stable gradient propagation in settings where standard variational circuits alone are known to exhibit barren plateau behavior.

\begingroup
\section{Scalability}
\label{scalability}
Barren plateaus are a known problem for trainability of QML architectures where the gradients of the trainable parameters vanish exponentially with number of qubits. In the original QResNet paper, specific constructions were found that eliminated the barren plateau~\cite{PhysRevApplied.23.044046}. Since we have modified the architecture, in this section we similarly identify how to construct a scalable model that will avoid exponentially vanishing gradients. While the experimental simulation in this work employ a fixed number of residual blocks (L = 5), understanding the scaling properties of the architecture is essential for guiding future implementations on real quantum devices.

First we expand the recursive relationship in Eq~\ref{eq:4}. We will have $m$ residual layers that consists of $l$ hardware efficient ansatz layers over $n$ qubits. We look at measuring the expectation value of $Z$ for the top qubit. We then take the gradient with respect to an arbitrary parameter $\vartheta_k$. The details are given in Appendix~\ref{appen-A}. We investigate when this quantity exhibits barren plateaus, i.e. when the variance is exponentially suppressed as number of qubits increase:

\begin{multline}
    \frac{\partial}{\partial \vartheta_k}tr(\mathcal{E}^m_l(\rho) Z_0) = \left(\frac{1}{\prod_{i=0}^{m-1}(1+\beta_i^2)} \right) \\\left( \beta_k^2 \frac{\partial}{\partial \vartheta_k} tr(W_k\rho_0W_k^\dagger)\right. \\+\sum_{i<j,i=k\lor j=k}\beta_i^2 \beta_j^2 \frac{\partial}{\partial \vartheta_k} tr(W_jW_i\rho_0 W_i^\dagger W_j^\dagger Z_0) +\ldots\\+\left.\left( \prod_{i=0}^{m-1}\beta_i^2 \right) \frac{\partial}{\partial \vartheta_k} tr\left((\prod W_i) \rho (\prod W_i^\dagger)Z_0\right) \right) . ~\label{eq:rhograd}
\end{multline}
where we have omitted the argument of $W_i(\vartheta_i)$ for ease of notation. This sum has terms consisting of all combinations of layers that include the $k$th layer.

Note that each of these terms are the same as a regular QVC model with ansatz $\prod^j W_i(\vartheta_i)$ where $j$ goes from one ResNet layer up to $m$ total layers. To analyse the scaling, we can use known results from QVC. In particular, note that if all the $W_l(\vartheta_l)$ have sufficient depth to be a 2-design, then all of the terms will be a QVC with a 2-design unitary which has guaranteed barren plateaus. However, there would be no point in implementing such a deep ansatz for each residual layer and the advantage is to use shallow ansatze for each residual layer with increased expressivity with the number of residual layers.

To that end, we note Cerezo et al.~\cite{Cerezo2021}, who show that with a local observable, an ansatz for $n$ qubits with:
\begin{itemize}
    \item $O(\log(n))$ depth avoids barren plateaus;
    \item $O(\text{poly}(\log(n)))$ depth is in an intermediate region;
    \item $O(\text{poly}(n))$ depth has barren plateaus.
\end{itemize}
This is supported by McClean et al.~\cite{McClean2018} who numerically showed a similar transition for a hardware efficient ansatz.

We now consider each term in Eq~\ref{eq:rhograd}. The first term is equivalent to a QVC measurement with one residual layer with $l_0$ layers and will not change with number of layers but has guaranteed lack of barren plateaus if $l_0 \in O(\log(n))$. The last term is a QVC measurement with all residual layers together, and will therefore have barren plateaus if the number of residual layers is $\sum l_i \in O(\text{poly}(n))$. The intermediate layers are combinatorial and include every combination of every number of residual layers and will transition from barren plateau free terms up to the maximum term, which depends on the total number of layers. In particular, we use only one hardware efficient layer which would be a depth of $O(1)$. The covariances similarly have barren plateaus if the terms have barren plateaus but they are also exactly zero between terms with barren plateaus and terms without. For the full calculation that considers the covariance of the terms in the series, see the appendix. 

We must consider scaling with number of residual layers. If we have total number of layers $m \in O(\text{poly}(n))$ for $n$ qubits, then even with constant depth in each layer $l \in O(1)$, then the final term in Eq~\ref{eq:rhograd} will be equivalent to a QVC with $O(\text{poly}(n))$ depth and will exhibit barren plateaus. There will be terms that do not have barren plateaus however we have to consider the coefficients as well. We only consider the gradient expression, the analysis of the total variance of the gradient including covariances are in the appendix.

For $\beta_i=0$, the $i$th layer will be skipped and we will effectively have a smaller number of layers $m$. For $\beta_i = \pm1$, we have an equal classical mixture of the state before and after each residual layer. In this case, we have the coefficient:
\begin{equation}
    \frac{1}{\prod_{i=0}^{m-1}(1+\beta_i^2)} = \frac{1}{2^{m-1}},
\end{equation}
with all other $\beta_i^2$ vanishing.

If $m \in O(\text{poly}(n))$, then this term will exponentially reduce. To counteract it, consider the total numbers of terms in the sum
\begin{equation}
    \sum_{k=0}^{m-1} \binom{m}{k} = 2^{m-1}
\end{equation}
which means, if they all had constant variance, it could cancel the exponential. Unfortunately, we have just proven that the larger terms will have $O(\text{poly}(n))$ scaling and will vanish due to barren plateaus. We have terms up to $k=K \in O(\text{poly}(\log(n)))$ that will not vanish. This means we only have:

\begin{equation}
    \sum_{k=0}^{K} \binom{m}{k} <<  2^{m-1},
\end{equation}
many terms which are subexponential in $m$. Therefore there is a barren plateau as we increase the number of layers.

Finally, we note that $O(log(n))$ depth circuits can often be efficiently classically simulable~\cite{PhysRevX.12.021021,bermejo2026quantum,angrisani2026simulating}. However, $O(\text{poly}(\log(n))$ depth circuits will be superpolynomial but subexponential. This is a known limitation of traditional QVC models and limits the expressivity of QVC.

To conclude, for a QResNet with $m$ residual layers of $l$ depth each for $n$ qubits, we can only avoid barren plateaus and simultaneously be difficult to simulate classically in the cases where $l \in O(\text{poly}(\log(n)))$ and $m \in O(\text{poly}(\log(n)))$. Fortunately, as discussed in the paper, QResNets allow for significant depth reduction for equivalent accuracy compared to QVC. This is because a QResNet without barren plateaus is effectively $\sum_{k=0}^{K} \binom{m}{k}$ many equivalent $O(\text{poly}(\log(n)))$ deep QVC circuits and, therefore, has much more expressivity without inducing barren plateaus. Furthermore, in this work, we have shown numerically for MNIST, CIFAR-2, and SARFish datasets in Table~\ref{tab:qvc_qResNet_comparison} that a QResNet is able to reach similar accuracies to QVC for binary classification with only 5 layers of depth 1, which is within the $O(\text{poly}(\log(n))$ regime for $10-12$ qubits, demonstrating increased expressivity while severely reducing total circuit depth and preventing barren plateaus compared to QVC.

\endgroup

\section{Adversarial Attack on QResNet}
\label{adv-attack}
Adversarial attack is a security concern for classical machine learning systems due to their vulnerability to data manipulations. While QML research is rapidly progressing in the recent years, vulnerability of QML models has been tested and benchmarked in the recent literature~\cite{lu2020quantum, ren2022experimental, gong2022universal, wendlinger2024comparative, anil2024generating, winderl2024quantum, guan2021robustness, liu2020vulnerability, west2023benchmarking, 41z8-d3h9}. In this work, we systematically tested the robustness of our proposed model against adversarial attack considering both white-box and black-box scenarios. In white-box scenarios, the adversary has full access to the QResNet architecture, including the variational parameters, residual strengths, and gradients of the loss with respect to the input. In black-box attacks, adversarial examples are generated using a classical neural network and transferred to QResNet without access to its internal parameters which is more close to real-world scenario.

We consider adversarial perturbations generated using the Fast Gradient Sign Method (FGSM), a first-order gradient-based attack widely adopted for benchmarking robustness in both classical and quantum classifiers. Given an input sample $\mathbf{x}$ and its true label $y$, FGSM constructs an adversarial example
\begin{equation}
\mathbf{x}_{\mathrm{adv}} = \mathbf{x} + \epsilon \, \mathrm{sign}\left( \nabla_{\mathbf{x}} \mathcal{L}(\mathbf{x}, y) \right),
\end{equation}
where $\mathcal{L}$ denotes the classification loss and $\epsilon$ controls the perturbation magnitude. The perturbations are constrained to remain imperceptible at the pixel level while maximally increasing the classification loss.

\begin{table*}[t]
\centering
\caption{Performance comparison between standard QVC architecture and the proposed QResNet models across multiple datasets. Total gate counts are computed for 10 data qubits. For QVC-200, the classification accuracy is reported from Ref. \cite{west2023benchmarking}. The gates counts are calculated from the circuit structure.}
\label{tab:qvc_qResNet_comparison}
\begin{tabular}{lccccccc}
\toprule
Model  & Qubits & Layers & Dataset & Task & Test Acc. (\%) & Total Gates \\
\midrule
QResNet (5 blocks)  & 10 & 5 & MNIST & Binary & 99 & 200 \\
QResNet (5 blocks)  & 10 & 5 & CIFAR-2 & Binary & 76 & 200 \\
QResNet (5 blocks)  & 10 & 5 & SARFish & Binary & 72.14 & 200 \\
\midrule
QVC-200 & 10 & 200 & MNIST & Multi-class (10) & 85 & 8000 \\
QVC-30 & 10 & 30 & MNIST & Multi-class (10) & 65 & 1200 \\
QVC-30 + QResNet  & 10 & 30 + 5 & MNIST & Multi-class (10) & 80 & 1400 \\
\bottomrule
\end{tabular}
\end{table*}

\section{Training Method}

All simulations were performed using the PennyLane framework \cite{bergholm2018pennylane} with PyTorch \cite{paszke2019pytorch} as the classical backend. The Adam \cite{kingma2014adam} optimizer was used with a learning rate of $5\times 10^{-3}$ and weight decay of $10^{-4}$. For binary classification tasks (digits \{0,1\} from MNIST, airplane and automobile from CIFAR-2, and fishing/not-fishing from SARFish), the model was trained with the binary cross-entropy loss, using 5 QResNet residual blocks, a batch size of 32, and 30 training epochs. For multi-class classification (digits \{0--9\} of MNIST), the circuit output was treated as a logit vector and trained with the standard cross-entropy loss. 
We found that a network with only 5 QResNet layers lacked sufficient expressivity for multi-class learning. 
To address this, we added 30 QVC layers before the residual blocks. Training was performed with a batch size of 256 for 5 epochs. In all cases, the trainable parameters included both the variational angles of the strongly entangling layers and the residual strengths $\beta_l$. All experimental simulations were conducted on a high-performance computing system equipped with a single NVIDIA GPU.

\section{Results and Analysis}
\label{results}
We benchmark our proposed method on MNIST~\cite{lu2020quantum} and CIFAR-2~\cite{krizhevsky2009learning} datasets and demonstrate practical utility using SARfish~\cite{luckett2024sarfish} dataset. Details of the datasets and experimental setup are provided in Appendix \ref{datasets}. The results are reported in Table \ref{tab:qvc_qResNet_comparison}.

We first evaluate the proposed QResNet framework on three benchmark binary classification tasks: MNIST (0,1), CIFAR-2 (airplane vs. automobile), and SARFish (fishing vs. non-fishing vessels). Table \ref{tab:qvc_qResNet_comparison} summarizes the test performance and gate count comparison across datasets. For MNIST, the model test accuracy is exceeding 99\% within the first few epochs while requiring only 200 quantum gates. QResNet performs learning directly within the quantum circuit through residual unitary transformations. CIFAR-2 results highlight the challenge of classifying the images with greater intra-class diversity and background complexity. While the training loss decreases more slowly than for MNIST (see Figure \ref{fig:mnist-binary-result}), the network nonetheless achieves robust generalization, with 76\% test accuracy using substantially fewer gates. The SARFish performance further demonstrate the practical applicability of the model to real-world remote sensing problems. Unlike MNIST, CIFAR, SAR data are noisy, sparse, and structurally distinct. Nevertheless, QResNet achieves consistent learning dynamics, with steady reduction of the training loss and the test accuracy is 72.14\%.

We also evaluate the proposed model on the full-scale 10-class MNIST dataset, and the quantitative performance comparison is summarized in Table I together with the corresponding quantum resource requirements. The results demonstrate clear differences between conventional variational quantum classifiers and residual quantum architectures under identical qubit configurations. A QVC-only baseline consisting of 30 variational layers achieves approximately 65\% classification accuracy, indicating limited representational capability when circuit depth is constrained for hardware feasibility. As shown in Table \ref{tab:qvc_qResNet_comparison}, adding the same 30-layer QVC backbone with only five ancilla-controlled QResNet residual blocks substantially improves classification performance, increasing the test accuracy to approximately 80\% while requiring only a moderate increase in total gate count. Notably, this improvement is obtained without increasing the variational circuit depth itself. The residual construction enables adaptive interpolation between identity evolution and parameterized unitary transformations, allowing quantum information to bypass non-essential operations during optimization. Consequently, the proposed architecture improves optimization stability and expressive capability while maintaining shallow circuit structure suitable for near-term quantum implementation. These results are consistent with the theoretical analysis in Section \ref{scalability}
, where QResNet constructions with shallow residual depth can retain expressive power without requiring the O(poly(n))-depth scaling typically associated with highly expressive QVC architectures. Despite employing only five residual blocks, the proposed model achieves performance comparable to substantially deeper QVC baselines, demonstrating that the residual formulation can improve representational capacity while maintaining hardware-efficient shallow circuits.

\begin{figure}
\centering
\includegraphics[width=1.0\linewidth]{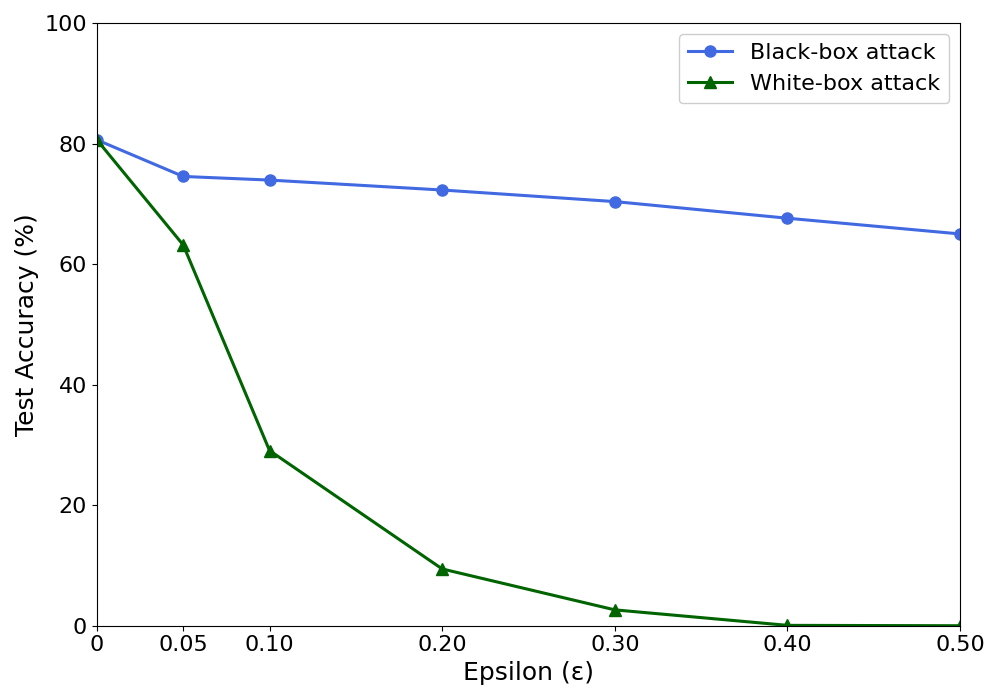}
   \caption{\textbf{Adversarial robustness of QResNet on the MNIST 10-class dataset.} The figure shows test accuracy as a function for both white-box and black-box attack settings. Under white-box attacks, classification accuracy decreases rapidly as $\epsilon$ increases. In black-box attacks, the QResNet model maintains substantially higher, demonstrating strong resilience to transferred adversarial examples from classical ML model.}
\label{fig:adv_accuracy}
\end{figure}

Table \ref{tab:qvc_qResNet_comparison} further highlights the significant reduction in quantum computational complexity achieved by QResNet compared with deep QVC architectures. For ten data qubits, a standard QVC comprising 200 variational layers \cite{west2023benchmarking} requires approximately 8000 logical quantum gates, including 6000 single-qubit rotations and 2000 entangling CNOT operations, to achieve nearly 85\% accuracy on the multi-class MNIST task. In contrast, our QResNet architecture employs only five residual blocks requiring 200 total gates, including 150 single-qubit rotations and 50 CNOT gates, representing a substantial reduction in circuit complexity. When combined with a shallow 30-layer QVC backbone, comparable classification performance is obtained using only 1400 gates, yielding a significant improvement in hardware efficiency relative to deep variational circuits. Since entangling operations constitute the dominant source of noise and decoherence in NISQ devices, reducing overall circuit depth directly improves practical implementability. The corresponding training loss evolution and convergence behavior for all evaluated architectures are provided in the Appendix for completeness. We additionally note that simulation of deeper QResNet configurations was limited by classical high-performance computing memory constraints arising from ancilla-controlled residual operations, which scale exponentially in state-vector simulation. On physical quantum hardware, where memory scales linearly with qubit count, deeper residual quantum architectures are expected to remain feasible.

Furthermore, we test the robustness of our model against adversarial attacks in both white-box and black-box scenarios using FGSM perturbations. The generated adversarial examples are shown in Figure \ref{fig:adv_examples} in the Appendix. The test accuracy is reported in Figure~\ref{fig:adv_accuracy}. While the model shows vulnerability under white-box attacks, where the adversary has full access to the quantum architecture, parameters, and gradients, the classification accuracy degrades gradually with increasing perturbation strength, as expected for fully differentiable models. In contrast, QResNet demonstrates strong robustness in the black-box setting, where adversarial examples generated from a classical neural network exhibit limited transferability to the quantum model. For multi-class MNIST tasks, the performance remains largely stable under black-box attacks even at higher perturbation magnitudes. This result indicates that the decision boundaries learned by the proposed method are structurally misaligned with those of classical models, thereby reducing the effectiveness of transferred adversarial perturbations.

\section{Conclusion}
In this work, we propose a QML approach that enables stable and fully differentiable training of deep quantum models without relying on post-selection and with significantly fewer gates required for high accuracy. By incorporating ancilla-controlled residual connections with trainable strengths, the proposed architecture addresses the trainability limitations of standard QVC and provides a mechanism for mitigating barren plateaus. Through systematic evaluation on image classification tasks, we demonstrate that introducing only a small number of quantum residual blocks leads to significant improvements in optimization stability and generalization performance for fewer overall gates. These performance gains might be due to the residual connections adaptively balancing identity and variational transformations, thereby preserving gradient flow during training. In addition, QResNet exhibits robustness to black-box adversarial attacks, suggesting that residual structure contributes to smoother decision boundaries and improved stability under adversarial perturbations. These findings establish quantum residual learning as a key architectural principle for scalable and robust QML, and provide a practical pathway toward deployable models on near-term quantum hardware.

\section*{Data Availability Statement}
All datasets generated in this work are available in Figures. Further information can be provided upon reasonable request to the corresponding author.

\section*{Acknowledgments}
Authors acknowledge the use of CSIRO HPC for conducting all the experimental simulations. We acknowledge Haiyue Kang for feedback and discussions. This activity is supported by the Advanced Strategic Capabilities Accelerator’s Emerging and Disruptive Technology program, delivered by the Defence Science and Technology Group (DSTG).

\bibliographystyle{unsrt} 
\bibliography{apssamp}

\clearpage
\appendix
\begingroup
\section{Scalability Calculation with Covariance Analysis}
\label{appen-A}
Let us have a Qresnet with $m$ layers where each residual layer is given by $U_i$ with coefficient $\beta_i$. The input state is $\rho_0$ and we measure the $Z$ expectation of the top qubit after $m$ layers so $tr(\rho_m Z_0)$. The gradient for a given $U$ selects out only the terms with $U_k$ and is given by:
\begin{multline}
     \frac{\partial tr(\rho_m Z_0)}{\partial U_k} = \frac{1}{\prod_{i=0}^{l-1}(1+\beta_i^2)} ( \beta_k^2 \frac{\partial}{\partial U_k} tr(U_k\rho_0U_k^\dagger)+ \\\sum_{i<j,i=k\lor j=k}\beta_i^2 \beta_j^2 \frac{\partial}{\partial U_k} tr(U_jU_i\rho_0 U_i^\dagger U_j^\dagger Z_0) +\ldots \\+\left( \prod_{i=0}^{m-1}\beta_i^2 \right) \frac{\partial}{\partial U_k} tr\left((\prod U_i) \rho (\prod U_i^\dagger)Z_0\right) ) ,
\end{multline}
and as the number of qubits increase, we increase the dimension of the unitaries $U_i$. First note that each $U_i$ consists of $l_i$ layers, so applying multiple unitaries is equivalent to increasing the number of layers. This means our overall function is simply the sum of terms with increasing numbers of layers with all combinations with correct ordering. We can therefore calculate the gradients of each term by considering the general case of an ansatz with $l_i$ layers. The gradient with respect to the parameter $\vartheta_\mu$ which controls a $e^{-i \vartheta_\mu H_\mu}$ term is given by:

\begin{equation}
    \frac{\partial}{\partial \vartheta_\mu} = i~tr(\rho U^\dagger_- [Z_0, U_+^\dagger H_\mu U_+]U_-),
\end{equation}
where $U_+$ are the layers after $H_\mu$ and $U_-$ are the layers before. It is shown by McClean et al.~\cite{McClean2018} that if either of these form a 2-design, then the overall term will have a barren plateau. Arbitrary depth ansatze for the residual layers will therefore result in barren plateaus.

However, it was shown~\cite{Cerezo2021} that an ansatz with number of layers $l \in O(\log(n))$ for $n$ qubits can be constructed such that the gradient
\begin{equation}
    \frac{\partial C}{\partial U_j} \ge G \in \Omega \left(\frac{1}{\text{poly}(n)} \right),
\end{equation}

\noindent with an intermediate region with $O(\text{poly}(\log(n)))$ layers to transition to the barren plateau depth.

This means that our overall function, since it contains all combinations of layers $\sum l_i$, will be the sum across the entire space from $O(\log(n))$ with guaranteed no barren plateaus to $O(\text{poly}(\log(n)))$ and eventually $O(\text{poly}(n))$ where there are guaranteed barren plateaus. 

To evaluate it explicitly, we first note that all these terms have expectation values of $0$ by construction of the ansatz. Explicitly:
\begin{equation}
    \mathbb{E}\left[\frac{\partial tr(\rho_n Z_0)}{\partial U_k} \right] = 0.
\end{equation}

\noindent Therefore, 
\begin{equation}
    \mathrm{Var}\left[\frac{\partial \mathrm{tr}(\rho_n Z_0)}{\partial U_k}\right] = \mathbb{E}\left[\left(\frac{\partial \mathrm{tr}(\rho_n Z_0)}{\partial U_k}\right)^2\right].
\end{equation}
This will simply be the multiplication of the sum by itself consisting of squared elements (variances) and cross terms (covariances). As discussed, this will have barren plateaus for $\sum l_i\in \text{poly}(n)$ and no barren plateaus for $\sum l_i\in \log(n)$. 

As for the covariances, they appear of the form:

\begin{equation}
    tr(\rho U^\dagger_- [Z_0, U_+^\dagger H_\mu U_+]U_-) tr(\rho U'^\dagger_- [Z_0, U_+'^\dagger H_\mu U'_+]U'_-),
\end{equation}
where $U'$ consists of layers that are a subset of the layers in $U$. 

The total variance for $m$ layers is therefore a sequence of $2^{2m}$ terms where $2^m$ are variances for each term and the rest are covariances. We will first investigate how increasing the qubit number will cause exponential concentration of each term, we will then see how increasing the layers can cause exponential concentration.

Firstly, note that the derivative over a parameter $\vartheta_\mu$ that is in $U$ but not $U'$ will result in the covariance to be zero. Therefore, training parameters on layers other than the initial residual layer will always have a set of vanishing covariances.

Secondly, we consider if $U_-$ forms a 2-design, i.e. there are $\text{poly}(n)$ layers in the smaller term before the angle $\vartheta_\mu$. In this case, we can write $U_- = W U_-' V$ and letting $A = [Z_0, U_+^\dagger H_\mu U_+]$, we can write:

\begin{multline}
    \mathbb{E}_{U_-}[tr(V\rho V^\dagger  U'^\dagger_- W^\dagger A WU'_-) tr(\rho U'^\dagger_- AU'_-)] = \\
    \frac{1}{d^2-1}(tr(A)tr(W^\dagger A W) tr(\rho)tr(V \rho V^\dagger) \\- \frac{1}{d}(tr(A)tr(W^\dagger A W) tr(\rho V \rho V^\dagger) \\+ tr(A W^\dagger A W) tr(\rho)tr(V \rho V^\dagger)) + tr(A W^\dagger A W) tr(\rho V \rho V^\dagger)),
\end{multline}
from Weingarten calculus. Noting that $tr(A) = 0$, we can simplify to:
\begin{equation}
    \frac{1}{d^2-1}\left( tr(A W^\dagger A W) \left(tr(\rho V \rho V^\dagger)- \frac{1}{d}(tr(\rho)tr(V \rho V^\dagger) \right) \right).
\end{equation}
If we assume $W$ has a 1-design, we can integrate over $W$ but $E_W[tr(AW^\dagger A W)] = \frac{1}{d} tr(A)^2 = 0$. Similarly, if we now assume $V$ has a 1-design, note that $E[tr(\rho)tr(V \rho V^\dagger)] = \frac{1}{d} tr(\rho)^2$ and $E[\frac{1}{d}(tr(\rho)tr(V \rho V^\dagger)] = \frac{1}{d}tr(\rho)^2$, so the overall expression becomes $0$. 

Even if neither $V$ nor $W$ are a 1-design, this expression has barren plateaus and further integration of $U_+$, $W$, or $V$ will not be able to mitigate them. 

Similarly, we can assume $U_+$ forms a 2-design. Note that we can rearrange the commutator since $\text{tr}(A[B,C]D) = tr([A,B]CD)$ due to cyclicity of the trace. This means 
\begin{equation}
    tr(\rho U^\dagger_- [Z_0, U_+^\dagger H_\mu U_+]U_-) =tr(H_\mu U_+[U_-\rho U^\dagger_- ,Z_0] U_+^\dagger ),
\end{equation}
and so we have the exact same results where either the covariances are zero or exhibit a barren plateau. 

We consider now the case where $U'$ consists of $\log(n)$ (or $\text{poly}\log(n)$) layers. We still split $U$ into $U_- = V U'_- W$. Firstly, if $U$ also has $\log(n)$ (or $\text{poly}\log(n)$) layers, this will be similar to the variance terms and will not exhibit barren plateaus. We do not integrate over $U'_-$ as before, but consider either $W$ or $V$ to form a 1-design. Integrating over $W$ gives:

\begin{multline}
   \mathbb{E}_W[tr(\rho V^\dagger  U'^\dagger_- W^\dagger A WU'_-V) tr(\rho U'^\dagger_- AU'_-)] \\= tr(\rho U'^\dagger_- AU'_-)E[tr(\rho V^\dagger  U'^\dagger_- W^\dagger A WU'_-V)] \\= \frac{1}{d} tr(A)tr(U_-'V \rho V^\dagger U_-'^\dagger) = 0,
\end{multline}
since $tr(A) = 0$. Similarly, assuming that $V$ is a 1-design and integrating over it first results in:

\begin{equation}
    \frac{1}{d} tr(\rho) tr(U'^\dagger_- W^\dagger A WU'_-) = \frac{1}{d}tr(\rho)tr(A) = 0.
\end{equation}

In summary, the variance of the gradient will result in the sum of variance of each individual term which will either exhibit no barren plateaus for $\sum l_i \in \log(n)$ depth or have guaranteed barren plateaus for $\sum l_i \in \text{poly}(n)$ depth. There will also be additional covariances which:
\begin{enumerate}
    \item exhibit no barren plateaus if all contributing terms have no barren plateaus;
    \item vanish (are exactly zero), in particular if one term has a barren plateau and the other does not;
    \item have barren plateaus.
\end{enumerate}
This means that, as we increase the number of qubits, our gradient will, firstly, always include individual trainable terms and, secondly, automatically reach the limit of trainability by considering every term and letting the gradients vanish for any term with barren plateaus.

For scaling with number of layers, we have calculated the variance and covariances of each term but now we have to consider the $\beta_i$ coefficients. If $\beta_i = 0$ or $\beta_i = \pm \infty$, then it will remove terms from the sum either being effectively fewer layers or an extra QVC layer respectively. The worst case for scaling will be if $\beta_i = \pm1$, where there is equal distribution of weights between applying and not applying the residual layer. In this case we have:
\begin{multline}
    \mathrm{Var}\left[\frac{\partial tr(\rho_n Z_0)}{\partial U_k} \right] = \\\frac{1}{2^{2(m-1)}} \left( \sum_i \mathrm{Var}[t_i] + \sum_{i,j} \mathrm{Covar}[t_i,t_j] \right),
\end{multline}
for each trace term $t_i$ in the original sum. These terms consist of all combinations of subsets of $m$ layers. This means that there are:
\begin{equation}
    \sum_{k=0}^{m-1} \binom{m}{k} = 2^{m-1}
\end{equation}
variance terms and subsequently $2^{2(m-1)}$ total terms. Assuming all variances and covariances are approximately equal, then they can cancel the pre-factor and have no concentration, but we have already shown above that this is never the case. In the general case, this will have exponential suppression with layers, i.e. a barren plateau. However, we will now investigate how to guarantee subexponential scaling.


Consider the best case scenario with $l_i = O(1)$ constant depth for each residual layer. In this case, arbitrary $m$ residual layers will result in terms from $\sum l_i \in O(\log(n))$ to $\sum l_i \in O(\text{poly}(n))$ and so on. The number of terms without barren plateaus will be:
\begin{equation}
    \sum_{k=0}^{K} \binom{m}{k},
\end{equation}
for $K \in O(\text{poly}\log(n))$. In the best case scenario their covariances exist and we have $\left( \sum_{k=0}^{K} \binom{m}{k} \right)^2$ total terms. The remaining terms are vanishing covariances between these terms and the $O(\text{poly}(n))$ scaling terms; or either terms with barren plateaus or vanishing covariances. Finally, note that $\sum_{k=0}^{K} \binom{m}{k}$ is significantly smaller than $2^{2(m-1)}$ and the vast majority of terms will have barren plateaus. If, however, we let $m \in O(\text{poly}\log(n))$, then the prefactor $\frac{1}{2^m}$ will decay subexponentially with $n$ and we will avoid a barren plateau. Note that this is still the case if $l_i \in O(\text{poly}(\log(n))$ since the maximum depth term is $\sum_i^m l_i \leq m\cdot|l|_{max} \in O(\text{poly}(\log(n))$.

In conclusion, this means that the only way to scale the QResNet is to have depth of each residual layer, $l_i \in O(\text{poly}\log(n))$ and the number of layers $m\in O(\text{poly}\log(n))$.

\endgroup


\begin{figure*}
\begin{center}
\includegraphics[width=1.0\linewidth]{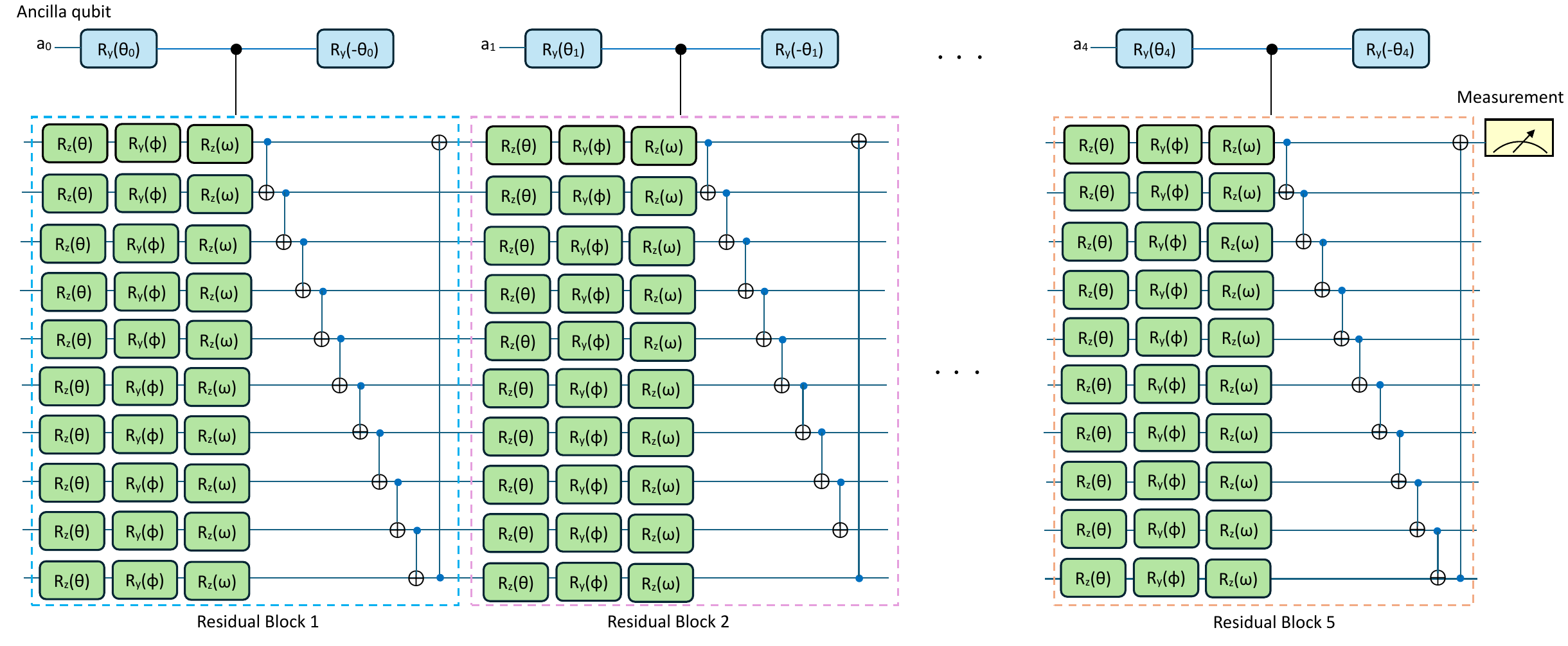}
\end{center}
   \caption{\textbf{Detailed circuit of the ancilla-controlled residual block in QResNet.}
Each residual block is controlled by an ancilla qubit $a_\ell$, which is first prepared in a superposition using a rotation $R_Y(\theta_\ell)$. 
This ancilla acts as the control for a parameterized variational unitary applied to the data qubits, implementing the residual transformation via an ancilla-controlled quantum channel. The ancilla qubit is subsequently unprepared using the inverse rotation $R_Y(-\theta_\ell)$. Within each residual block, parameterized single-qubit rotations $R_z(\theta)$, $R_y(\phi)$, and $R_z(\omega)$ are applied to each data qubit, followed by entangling operations between neighboring qubits. 
Multiple residual blocks are applied sequentially, with each block controlled by a different ancilla qubit. 
After the final residual block, measurements are performed on the data qubits to obtain expectation values used for downstream classification.}
\label{fig:circuit}
\end{figure*}

\begin{figure*}
\begin{center}
\includegraphics[width=0.98\linewidth]{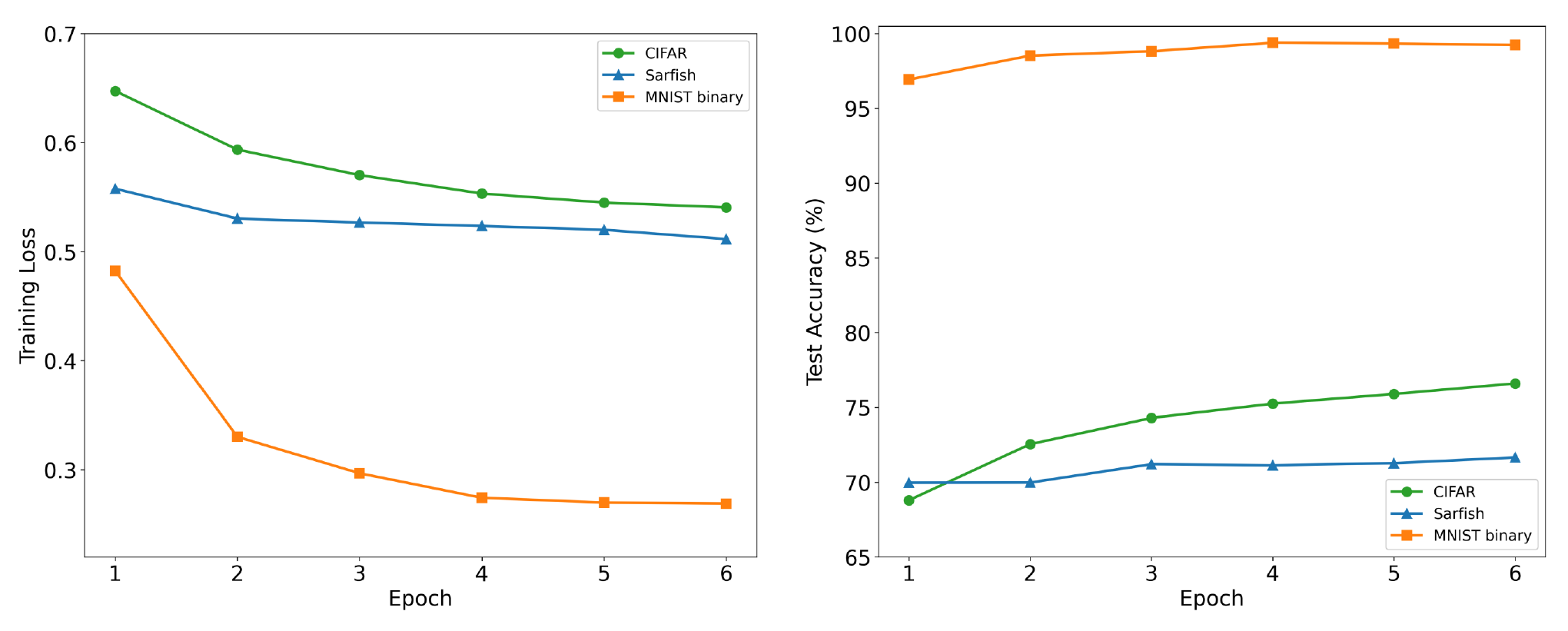}
\end{center}
   \caption{\textbf{Training dynamics and generalization performance of the QResNet architecture on binary classification tasks.}
The left panel reports the training loss as a function of epoch, while the right panel shows the corresponding test accuracy. Results are presented for three binary datasets: MNIST (digits 0 vs 1), CIFAR-2 (airplane vs. automobile), and SARFish (fishing vs. non-fishing vessels).}
\label{fig:mnist-binary-result}
\end{figure*}

\begin{figure*}
\begin{center}
\includegraphics[width=0.98\linewidth]{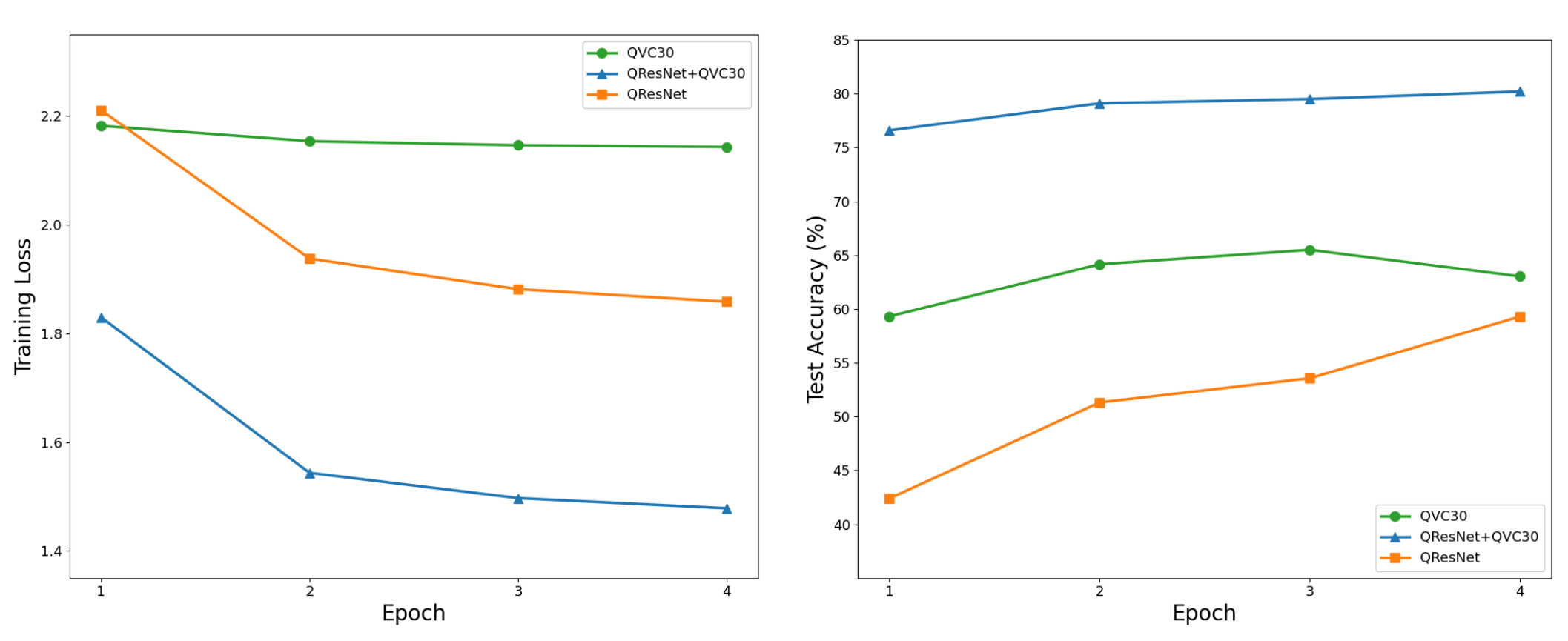}
\end{center}
   \caption{\textbf{Multi-class MNIST classification:} Training loss and test accuracy for 10-class MNIST classification comparing QResNet with baseline QVC models. Panels show the training loss and test accuracy, respectively, for three architectures: a QVC-only model with 30 layers (QVC-30), a model with 30 QVC layers followed by QResNet residual blocks (QResNet + QVC-30), and the QResNet architecture with 5 layers only (no QVC). The results illustrates that the classification performance drops from 80\% to 65\% when there is no residual connection.}
\label{fig:mnist-multi-result}
\end{figure*}

\begin{figure*}
\begin{center}
\includegraphics[width=1.0\linewidth]{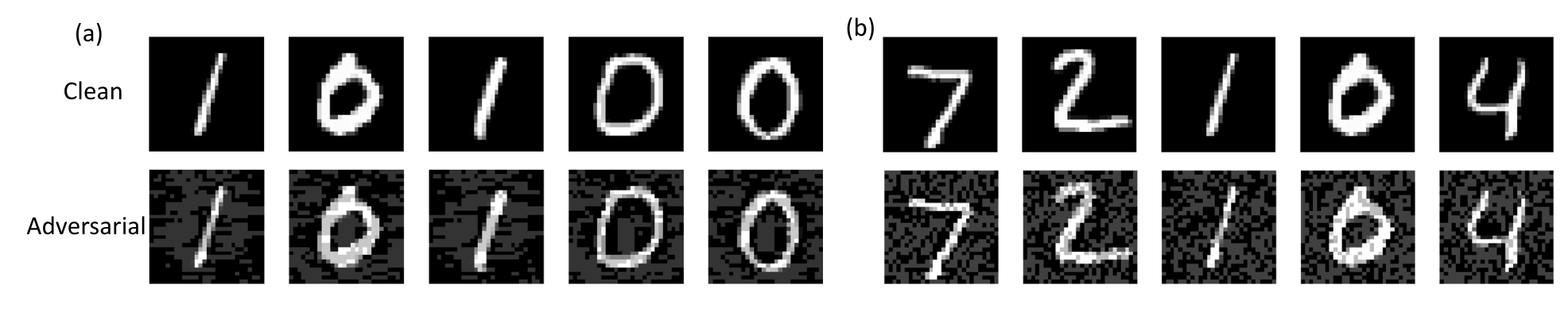}
\end{center}
   \caption{\textbf{Adversarial attack on MNIST binary and multi-class datasets.} The top row shows the clean images, and the bottom one represents the adversarial images generated by FGSM adversarial attack.}
\label{fig:adv_examples}
\end{figure*}

\section{Datasets}
\label{datasets}
In this section, we describe the datasets used to evaluate the performance of the proposed model across both binary and multi-class classification settings.

\textbf{MNIST}~\cite{lu2020quantum} is one of the most widely used benchmarks in ML. It contains 70,000 grayscale images of handwritten digits from 0 to 9. Each image has a resolution of $28 \times 28$ pixels, which corresponds to 784 features. The dataset is divided into 60,000 training images and 10,000 test images, with an equal number of samples from each digit class. In our experiments, we tested our model on two settings. For binary classification, we selected the digits {0,1}. For multi-class classification, we used all ten digit classes.

\textbf{CIFAR-10}~\cite{krizhevsky2009learning} (Canadian Institute for Advanced Research-10) is another popular benchmark for image classification. It contains 60,000 color images belonging to 10 different object categories. Each image has a resolution of  $32 \times 32$ pixels and three color channels (red, green, and blue). The standard split consists of 50,000 training images and 10,000 test images, with the same number of samples per category. For our experimental simulation, we focused on a binary subset of CIFAR-10, which we refer to as CIFAR-2, containing the classes airplane and automobile. This subset contains 10,000 training images and 2,000 test images, equally divided between the two categories. CIFAR-2 is more challenging than MNIST dataset because the images have higher dimensionality and include more variability in textures, colors, and backgrounds.

\textbf{SARfish}~\cite{luckett2024sarfish} is a dataset to identify ships using Synthetic Aperture Radar (SAR) data collected along a coastline with corresponding xView3 labels. The identification of ships by SAR data can aid in monitoring, control, and surveillance of illegal, unreported, and unregulated fishing activity. If left unchecked, such fishing activity can disrupt the natural ecosystem and lead to overfishing, which will impact marine biodiversity and limit food security for the reliant communities. 

\section{Training dynamics and generalization performance of the QResNet }
Figure \ref{fig:mnist-binary-result} summarizes the training loss and test performance across datasets. For MNIST, the model exhibits faster convergence, with the training loss decreasing with the number of epochs and the test accuracy is exceeding 99\% within the first few epochs. CIFAR-2 results highlight the challenge of classifying the images with greater intra-class diversity and background complexity. While the training loss decreases more slowly compared to MNIST, the network nonetheless achieves robust generalization, with 76\% test accuracy. The SARFish performance further demonstrate the practical applicability of the model to real-world remote sensing problems. Unlike MNIST, CIFAR, SAR data are noisy, sparse, and structurally distinct. Nevertheless, QResNet achieves consistent learning dynamics, with steady reduction of the training loss and the test accuracy is 72.14\%.

We also report the proposed model on the full-scale 10-class MNIST dataset, as shown in Figure \ref{fig:mnist-multi-result} where we report the training loss and test accuracy for three architectures: a QVC-only baseline with 30 variational layers, a model comprising 30 QVC layers followed by 5 ancilla-controlled QResNet residual blocks, and only 5 ancilla-controlled QResNet when there are no QVC layers. The QVC-only baseline exhibits slower convergence and limited generalization, with the test accuracy saturating at approximately 65\%. In contrast, adding the same 30-layer QVC backbone with only 5 QResNet residual blocks leads to a substantial improvement in both optimization stability and classification performance, increasing the test accuracy to approximately 80\%. Importantly, this gain is achieved without increasing the depth of the variational circuit. Through trainable residual strengths, the QResNet layers enable adaptive interpolation between identity and variational transformations, allowing information to bypass non-essential operations when beneficial. This mechanism stabilizes optimization, mitigates gradient degradation, and significantly enhances expressivity in high-dimensional classification tasks, even when only a small number of residual layers are introduced.



\end{document}